\documentclass[12pt]{article}
\usepackage{epsf}
\usepackage{amsmath}

\usepackage{graphics}
\usepackage{cite}

\begin{titlepage}

\begin{document}

\hfill January 2021\\

\hfill Revised and Updated\\

\begin{center}

{\bf \Large Refined Mass Estimate for Bilepton Gauge Boson\\
}
\vspace{2.5cm}
{\bf Claudio Corian\`{o}\footnote{claudio.coriano@le.infn.it} and Paul H. Frampton\footnote{ paul.h.frampton@gmail.com} \\  }
\vspace{0.5cm}
{\it Dipartimento di Matematica e Fisica "Ennio De Giorgi",\\ 
Universit\`{a} del Salento and INFN-Lecce,\\ Via Arnesano, 73100 Lecce, Italy
}

\vspace{1.0in}

\begin{abstract}
\noindent
Using the most recent experimental data on parameters of the standard electroweak
theory, as well as renormalisation group equations with a boundary matching condition, we derive a
refined and more accurate value for the mass of the doubly-charged bilepton 
($Y^{\pm\pm}$) occurring in the spontaneous breaking of the gauge group
$SU(3)_L \times U(1)_X$ to the standard electroweak gauge group
$SU(2)_L \times U(1)_Y$.  Our result is $M(Y^{\pm\pm}) = (1.29 \pm 0.06)$ TeV.

\end{abstract}

\end{center}
\end{titlepage}

\noindent
A particle theory model which explains the occurrence of 
three quark-lepton families in the standard model is the bilepton model 
with gauge group $SU(3)_C \times SU(3)_L
\times U(1)_X$. For the minimal 331-model with no additional leptons, 
as proposed in \cite{PHF1992}, this involves, {\it inter alia}, two additional gauge bosons
$Y^{\pm\pm}$ with double
lepton number $|L|=2$. The present brief article derives the $Y^{\pm\pm}$ mass
with new accuracy by using renormalisation group equations with
a boundary matching condition and is intended to assist ongoing searches at the Large Hadron Collider.

\bigskip

\noindent
We recall that the most important experimental discovery in particle physics during the
previous decade was that of the Higgs boson (H) \cite{Higgs,Higgs2} and subsequent partial
confirmation of the BEH mechanism \cite{BE,H} in the couplings of $H$
to the quarks and charged lepton of the third family, as well as to the electroweak
gauge bosons. Before the $H$ discovery
the Higgs mass, $M(H)$, was not sharply predicted by the standard model but
merely expected to be within the wide range
\begin{equation}
114 ~ {\rm GeV} < M(H) < 180 ~ {\rm GeV}
\label{Hmass}
\end{equation}
where the lower limit came from LEP experiments and the upper limit
came from the analysis of radiative corrections. The correct mass is 
now known to be \cite{Higgs,Higgs2} $M(H)\simeq125.3$~GeV.

\bigskip

\noindent
It is the singly-charged  weak intermediate gauge boson in the 
standard electroweak theory which is the
particle most closely similar to the doubly-charged bilepton  gauge boson.
In the pattern of spontaneous symmetry breaking in cooling from high temperature,
regarded like a family tree, the $Y^{\pm\pm}$ and $W^{\pm}$ are siblings.
According to the BEH mechanism, the mass of $W^{\pm}$ is given by
\begin{equation}
M(W^{\pm}) = \frac{1}{2} g_2 v
\label{Wmass}
\end{equation}
where $g_2$ is the $SU(2)_L$ gauge coupling constant and $v = <H>$ is the vacuum
expectation value of the Higgs field. Eq.(\ref{Wmass}) holds at tree level, and for the physical
renormalized values of the three gauge-invariant qualities involved.
From the 2020 Particle Data Group\cite{PDG}
tabulations we find that $M(W^{\pm}) \simeq 80.4$ GeV, $v \simeq 246$ GeV and
$g_2 \simeq 0.65$. 

\bigskip

\noindent
The goal of this paper is to show that, using the current experimental data on the parameters of the electroweak theory and renormalisation group equations with a boundary condition at the
$331 \rightarrow 321$ gauge symmetry breaking, it is possible to derive a more accurate value for the mass of the doubly charged gauge bosons of the bilepton model, which is of interest for the ongoing experimental searches.

\bigskip

\noindent
From the 2020 Particle Data Group\cite{PDG}, we find $\alpha_{em}^{-1}(M_Z) = 128.0$,
${\rm sin}^2\theta_{ew} (M_Z)= 0.231$ and therefore ${\rm cos}^2\theta_{ew} (M_Z) =0.769$.
It follows that
\begin{equation}
\alpha_Y^{-1}(M_Z)= \alpha_{em}^{-1}(M_Z) ~  {\rm cos}^2\theta_{ew} (M_Z) = 98.4
\label{alphaY}
\end{equation}

\noindent
and

\begin{equation}
\alpha_2^{-1}(M_Z)= \alpha_{em}^{-1}(M_Z) ~ {\rm sin}^2 \theta_{ew}(M_Z) = 29.5
\label{alpha2}
\end{equation}

\bigskip

\noindent
The bilepton mass arises from a BEH mechanism, similarly to Eq.(\ref{Wmass}),
\begin{equation}
M(Y^{\pm\pm}) = \frac{1}{2} g_3 V
\label{Ymass}
\end{equation}
where $g_3$ is the $SU(3)_L$ gauge coupling constant, equal to $g_2$ at the
symmetry breaking scale $V$ which is the vacuum expectation value $V=<\Phi>$
of the scalar field $\Phi$ which breaks $SU(3)_L$ to $SU(2)_L$.  Eq.(\ref{Ymass})
is strictly analogous to Eq.(\ref{Wmass}) in that it holds not only at tree level
but also for the renormalised physical values of all the three gauge invariant
quantities involved.

\bigskip

\noindent
In approaching Eq,(\ref{Ymass}), we might initially expect the uncertainty in the bilepton mass
to be as fractionally significant as for the Higgs mass in Eq.(\ref{Hmass}). Fortunately,
this is not the case because a matching boundary condition is provided \cite{PHF1992} by the
gauge group embedding $321 \subset 331$, where $3n1 \equiv SU(3)_C\times SU(n)_L\times U(1)$,
in the
renormalisation group running of the electroweak mixing angle ${\rm sin}^2\theta_{ew}(\mu)$.

\bigskip

\noindent
To discuss this boundary matching condition, we begin with the definition
\begin{equation}
{\rm sin}^2 \theta_{ew}(\mu) = \frac{\alpha_Y(\mu)}{
\alpha_2(\mu) + \alpha_Y(\mu)}
\label{sinesquared}
\end{equation}
and consider the renormalisation group equations, with $y= {\rm ln} (\mu/M_Z)$,
\begin{equation}
\alpha_Y^{-1}(\mu) = \alpha_Y^{-1}(M_Z) - \frac{41}{12 \pi}  y
\label{Y}
\end{equation}

\noindent
and

\begin{equation}
\alpha_2^{-1}(\mu) = \alpha_2^{-1}(M_Z) + \frac{19}{12 \pi}  y
\label{2}
\end{equation}

\bigskip

\noindent
The embedding of $SU(2)_L$ in $SU(3)_L$ provides an upper limit on $V$
through the requirement that
\begin{equation}
{\rm sin}^2 \theta_{ew} (V) = \frac{1}{4}
\label{quarter}
\end{equation}
which corresponds in Eq.(\ref{sinesquared}) to
\begin{equation}
\alpha_Y^{-1}(V) = 3\alpha_2^{-1}(V)
\label{three}
\end{equation}
which, with Eqs.(\ref{Y},\ref{2}), yields the value 
\begin{equation}
y = {\rm ln} \left(\frac{V}{M_Z} \right) = 3.81 ~~~ {\rm or} ~~~ V \simeq 4.1 ~  {\rm TeV}
\label{V}
\end{equation}

\bigskip

\noindent
Finally for the bilepton mass in Eq.(\ref{Ymass}) we need to use $g_3=g_2$ at
$\mu=V$ and Eq.(\ref{2}) to find
\begin{equation}
g_3 = 0.63
\label{g3}
\end{equation}
and hence, allowing a $5\%$ error,
\begin{equation}
M(Y^{\pm\pm}) = (1.29 \pm 0.06) ~ {\rm TeV}
\label{MassY}
\end{equation}

\bigskip

\noindent
This is a sharper prediction than any previous estimate because
of the use of Eq.(\ref{quarter}) to set a mass scale in the renormalisation
group equations. The result Eq.(\ref{MassY}) is good news for
any bilepton search in the Run 2 data from LHC, already in the cloud,
because the mass value is sufficiently small that there could be
sufficient events to find a signal and make a discovery.

\bigskip

\noindent
In principle the scalar VEV could be slightly smaller than the value
derived in Eqs. (\ref{quarter},\ref{three},\ref{V}) but that would imply
that the bilepton mass is slightly smaller than in Eq.(\ref{MassY})
and therefore that the probability of finding a signal would be
 slightly bigger. We hope that this refined prediction may help the experimental searches for the
bilepton from the available experimental data.

\bigskip

\noindent
We add some comments on the previous literature. There are at least two papers\cite{DMP,FP}
which discuss a non-existent "Landau pole" in the 
bilepton model \cite{PHF1992} at a scale $\mu\simeq4.1$GeV. This is a misunderstanding
about the special status of the 4.1TeV scale. There is no Landau pole, nor
does $g_X$ diverge there. When we rewrite the matching equation in the equaivalent form
\begin{equation}
\frac{g_X^2}{g^2} = \frac{s_W^2}{1 - 4s_W^2}
\label{signchange}
\end{equation}
the RHS has a pole at $s_Z^2=1/4$ but, on the LHS, $g_X$ is {\it finite} while $g^2=0$.
In the bilepton model, it is the embedding of $SU(2)_L$ into
$SU(3)_L$ which underlies the feature at 4.1GeV.  The point is that
the electroweak gauge group $SU(2)_L \times U(1)_Y$ can be embedded entirely
into $SU(3)_L$ if and only if $s_W^2=1/4$ exactly. For $s_W^2 < 1/4$ it 
requires $g^2 > 0$ while for $s_W^2>1/4$ it would require
$g^2 < 0$ which would violate unitarity. The zero in the denominator on the
LHS of Eq.(\ref{signchange}) reflects the changing sign of $g^2$. 
This is interesting, and somewhat unusual amongst BSM theories, in that the new physics 
including the bilepton must lie below 4.1 TeV.
"Landau pole" should refer to the discovery made by Landau in 1954 that
the coupling in QED diverges at an extremely high energy.
Landau concluded that QFT must be wrong but his objection was too hasty as it
was answered by the discovery of asymptotic freedom in 1973. 
Landau's "pole" has no connection to the bilepton mass.

\bigskip

\noindent
Concerning \cite{PP}, it
presents a model similar to \cite{PHF1992} but deviates 
in three places, all of importance. First, in their abstract
they state that lepton number is violated. Actually, it is conserved. Second, although
they assign correctly two quark families to triplets and one quark family
to an antitriplet, to cancel anomalies, they choose (u,d) rather than (t,b) to assign
to the antitriplet. This assignment leads, however, to strong disagreement 
with FCNCs as shown in \cite{Ng}.
The same disfavoured quark assignment was later made in \cite{JJ}.
The third and most important deviation of \cite{PP} from \cite{PHF1992} is that
in the text it predicts "new physics at an, in principle, arbitrary mass scale", 
overlooking the upper mass bound on new physics first noted in \cite{PHF1992}. We regard this
upper mass bound as
more important even than the explanation for three families.

 \bigskip

\noindent
Finally, the paper \cite{SSV} presented what
can be called a 331-model, a name later introduced in \cite{PHF1992}
for the bilepton model because of its gauge group. It uses, however, different fermion
representations such that no bilepton is present.

\end{document}